\documentclass[submission, Phys]{SciPost}
\usepackage{bm,latexsym,mathrsfs,enumerate,amssymb}
\usepackage{color}
\usepackage{appendix}
\usepackage{mathtools} 
\usepackage{makecell}

\usepackage[cal=boondoxo,frak=boondox]{mathalfa}

\hypersetup{breaklinks=true,unicode=true,urlcolor = blue,colorlinks = true,citecolor = blue,linkcolor = red!75!black}
\graphicspath{{./figs/}}
\renewcommand{\vec}[1]{\bm{#1}}
\definecolor{RoyalBlue}{rgb}{0.25,0.41,0.88}
\usepackage{savesym}
\usepackage[most]{tcolorbox}
\savesymbol{comment}
\begin{document}
\begin{center}{
\Large \textbf{
Curvature induced magnonic crystal in nanowires
}}\end{center}

\begin{center}
Anastasiia Korniienko\textsuperscript{1,2},
Volodymyr P. Kravchuk\textsuperscript{2,3*},
Oleksandr V. Pylypovskyi\textsuperscript{1},
Denis~D. Sheka\textsuperscript{1},
Jeroen~van~den~Brink\textsuperscript{2,4,5},
Yuri Gaididei\textsuperscript{3}
\end{center}

\begin{center}
{\bf 1} Taras Shevchenko National University of Kyiv, 01601 Kyiv, Ukraine
\\
{\bf 2} Leibniz-Institut f{\"u}r Festk{\"o}rper- und Werkstoffforschung, IFW Dresden, 01069 Dresden, Germany
\\
{\bf 3} Bogolyubov Institute for Theoretical Physics of National Academy of Sciences of Ukraine, 03143 Kyiv, Ukraine
\\
{\bf 4} Institute for Theoretical Physics, TU Dresden, 01069 Dresden, Germany
\\
{\bf 5} Department of Physics, Washington University, St. Louis, MO 63130, USA

* vkravchuk@bitp.kiev.ua
\end{center}

\begin{center}
\today
\end{center}


\section*{Abstract}
{\bf
A new type of magnonic crystals, curvature induced ones, is realized in ferromagnetic nanowires with periodically deformed shape. A magnon band structure of such crystal is fully determined by its curvature: the developed theory is well confirmed by simulations. An application to nanoscale spintronic devises with the geometrically tunable parameters is proposed, namely, to filter elements.   }

\vspace{10pt}
\noindent\rule{\textwidth}{1pt}
\tableofcontents\thispagestyle{fancy}
\noindent\rule{\textwidth}{1pt}
\vspace{10pt}

%
\section{Introduction}
\label{sec:intro}

Recent advance in fabrication of sub-micrometer sized magnetic nanowires of controlled shape \cite{Streubel16a,Gibbs14,May19} makes them promising construct elements for variety of spintronic devices. Curvilinear geometry opens new possibilities for the device miniaturization \cite{Fernandez17} and such curvilinear elements were suggested for using in computer memory and logic \cite{Parkin15,Hrkac11}, in magnon waveguides \cite{Chumak15}. That is why a clear understanding of physical picture of influence of the wire curvature on the magnetization dynamics is an actual problem of the modern micromagnetism. 

Probably the most prominent manifestation of the curvature in magnetic systems is a magnetochiral effect \cite{Hertel13a}. In thin curved wires, the chiral symmetry breaking results in emergent geometry-induced Dzyaloshinskii--Moriya interaction (DMI), for a review see \cite{Streubel16a}. A number of the magnetochiral effects in dynamics of the nanowire magnetization were recently predicted. 
Spin waves can be bound by a local bending of the wire \cite{Gaididei18a}. A wire twisting results in non-reciprocal spin-wave propagation \cite{Sheka15c}. Numerous curvature effects in the domain wall dynamics were found, namely the localized curvature defect (e.g. a bend of the wire) creates a pinning potential for the domain wall \cite{Yershov15b}; the curvature gradient acts as a driving force for the domain wall resulting in a fast translational motion free of the Walker limit \cite{Yershov18a}; torsion of the helix shaped wire can result in a negative mobility of the spin-torque driven domain wall \cite{Yershov16}, etc. These and others curvature induced phenomena may be explained within a general framework \cite{Sheka15}, which introduces a number of emergent geometry-induced  interactions. The latter are effectively generated by the common energy terms comprising spatial derivatives, e.g. in presence of the curvature the exchange interaction generates the effective DMI and anisotropy \cite{Gaididei14,Sheka15}, the intrinsic DMI generates effective anisotropy \cite{Kravchuk16a,Volkov18}. 

Here we consider the magnon spectrum of a wire with periodically deformed shape. We demonstrate that such geometry gives rise to the periodical potentials in equations for the magnon propagation, as a consequence, the band structure appears. The analysis is based on the previously developed approach \cite{Sheka15} for curvilinear ferromagnetic wires. The full scale numerical analysis of the band structure is supplemented by the analytically obtained formulae valid for the limit cases of small and large curvatures. Additionally, we use the direct simulations of the Landau--Lifshitz equation in order (i) to verify the obtained results and (ii) to demonstrate that the periodically deformed wire can be utilized as a magnon filter. The latter is an important issue for the spintronic devices relied on the magnon crystals \cite{Krawczyk14,Langner18,Gulyaev03,Chumak15,Chumak09a,Chumak08,Wang09a,Beginin18}.

A large area of modern spintronics deals with the data transfer and processing  based on the magnon waves propagation\cite{Chumak15}. In this case, magnonic crystals  \cite{Krawczyk14} are the key elements of the magnonic spintronic devices including resonators, generators, filters, wave-guides, see reviews Ref.~\cite{Krawczyk14,Chumak15,Lenk11} for more information. Magnonic crystal is a magnetic system with artificially introduced periodicity. The latter can be created by means of periodical arrangement of several different magnetic materials \cite{Krawczyk14,Wang09a,Gallardo19} or by using one material with periodically modulated geometrical parameters \cite{Krawczyk14,Chumak09,Beginin18} including arrays of interacting magnetic strips \cite{Krawczyk14,Gubbiotti10,Topp10} or dots \cite{Krawczyk14,Lenk11}. The periodically arranged magnetization patterns can be also considered as a magnon crystal, e.g. the DMI induced helical structure \cite{Garst17a,Kugler15} or skyrmion lattice \cite{Garst17a,Schwarze15}.

Magnonic crystal, which we propose here, can be thought as a combination of the last two methods. On the one hand, we consider a curvilinear nanowire made of a single material. On the other hand, the periodically modulated curvature induces the periodical magnetization pattern, see Figs.~\ref{fig1}, \ref{fig2}, resulting in periodical potentials. The magnon scattering on the periodical potentials induces the band structure. This process is completely analogous to formation of the energy zones in solid state crystals \cite{Kittel05}, but instead of the common Schr{\"o}dinger equation we have its generalized form see Sec.~\ref{sec:BS} for details.

All analytical predictions are in good agreement with the spin lattice simulations, see App.~\ref{sec:simuls}. In order to demonstrate that the meander-like periodical structure, see Fig.~\ref{fig1}(a), can be used as a magnon crystal we consider a 14-periods part of such structure as a magnon filter. Using the spin-lattice simulations based on the Landau--Lifshitz equation we obtain the amplitude-frequency characteristic. The input signal was formed by a homogeneous rotation of the first spin with the input frequency. The output signal demonstrates the presence of the gaps in the system. Therefore the corresponding frequencies are effectively filtered. Position and width of the band gaps are determined by the curvature of the wire, see  App.~\ref{sec:AFC}, Fig.~\ref{fig1}(d) and Supplemental movie~\footnote{\label{note1}See Supplemental movie at \texttt{Link provided by the publisher} of spin-lattice simulation of Landau--Lifshitz equations.} for details.

%
%
\section{Model and the equilibrium magnetization pattern}
\label{sec:Model}

\begin{figure}
	\includegraphics[width=\textwidth]{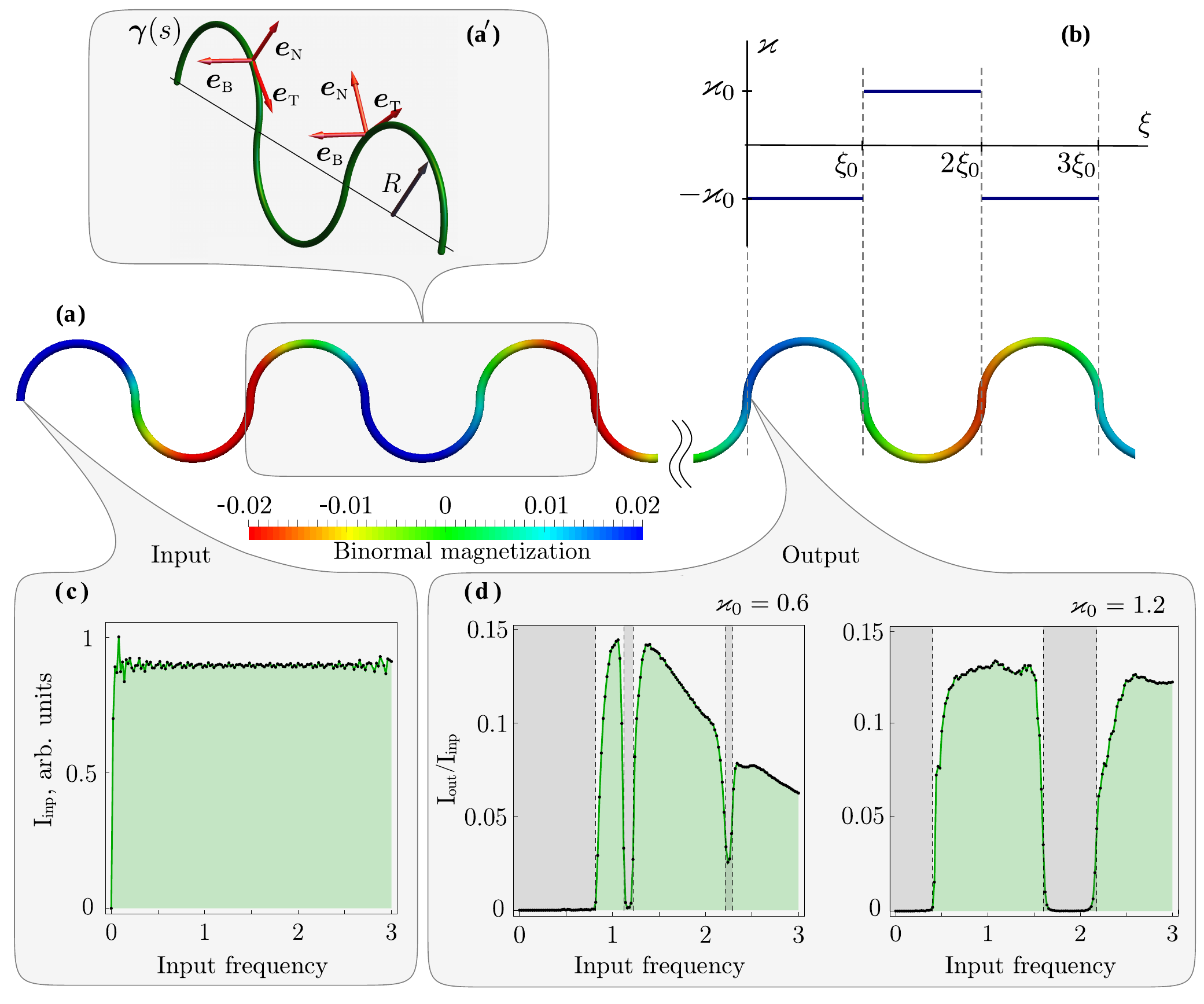}
	\caption{\textbf{Meander-like shape nanowire:} Wire in form of periodically repeated semicircles (a),(a$'$) has the square-wave curvature $\varkappa(\xi)$ (b) with dimensionless amplitude $\varkappa_0=\ell/R$ and period $2\xi_0=2\pi/\varkappa_0$. We demonstrate that the considered meander-like shape structure can be used as a magnon filter element using spin-lattice simulation based on Landau--Lifshitz equation with damping, for details see Appendix~\ref{sec:AFC}. Colour scheme shows the binormal magnetization distribution in the wire (a) with curvature $\varkappa_0 = 1.2$ and input signal of dimensionless frequency $\Omega =1$. Amplitude-frequency characteristics of input (c) and output (d) signals of a magnon filter element is composed of 14 periods of the meander shaped structure. The frequency ranges, where the magnons propagation is strongly depressed are shown by the grey shading. The band gaps position and width can be controlled by the parameter $\varkappa_0$. Decreasing the output amplitude with frequency takes place due to the natural damping ($\alpha=0.01$) included into the simulations. The terminator is placed after a magnon filter element to prevent signal reflection from the free boundary of the system. The terminator is composed of 5 periods of the meander shaped structure and has a spatially increasing damping coefficient, see Appendix~\ref{sec:AFC} for details. The magnetization dynamics are illustrated at the Supplemental movie~\textsuperscript{\ref{note1}}.}\label{fig1}
\end{figure}

A thin wire can be defined as a space domain  $\vec{r}(s,\xi_1,\xi_2)=\vec{\gamma}(s)+\xi_1\vec{e}_\textsc{n}(s)+\xi_2\vec{e}_\textsc{b}(s)$. Here $\vec{\gamma}(s)$ is the central line of the wire parametrized by its arc length $s$, coordinates $\xi_1$ and $\xi_2$ parametrize the perpendicular cross-section of the wire, vectors $\vec{e}_\textsc{n}$ and $\vec{e}_\textsc{b}$ are the normal and binormal Frenet–-Serret orts \cite{Kuehnel15} of the line $\vec{\gamma}$, respectively, see Fig.~\ref{fig1}(a$'$). The term `thin wire' means the simultaneous fulfilment of two conditions: geometrical one and magnetic one. The geometrical condition reads $|\partial_s^2\vec{\gamma}|\ll1/h_{\rm max}$, where $h_{\rm max}$ is the maximal lateral size of the perpendicular wire cross-section. The magnetic condition will be discussed latter. Together with the tangential unit vector $\vec{e}_{\textsc{t}}=\partial_s\vec{\gamma}$ the normal and binormal vectors compose the TNB basis $\{\vec{e}_{\textsc{t}},\vec{e}_{\textsc{n}},\vec{e}_{\textsc{b}}\}$. For the particular case of a planar wire lying within the $xy$-plane the TNB basis vectors can be presented in the form $\vec{e}_\textsc{t}(s)=\hat{\vec{x}}\cos\alpha(s)+\hat{\vec{y}}\sin\alpha(s)$, $\vec{e}_\textsc{n}(s)=-\hat{\vec{x}}\sin\alpha(s)+\hat{\vec{y}}\cos\alpha(s)$, and $\vec{e}_\textsc{b}=\hat{\vec{z}}$ with alternating-sign curvature $\kappa=\partial_s\alpha$.

Dynamics of the wire magnetization is governed by Landau--Lifshitz equation 
\begin{subequations}\label{eq:Model}
\begin{align}\label{eq:LL}
\partial_t\vec{m}=\frac{\gamma_0}{M_s}\vec{m}\times\frac{\delta E}{\delta\vec{m}},
\end{align}
where $\vec{m}=\vec{M}/M_s$ is the unit magnetization vector normalized by the saturation magnetization, $\gamma_0>0$ is the gyromagnetic ratio, $E$ is a total magnetic energy, and the damping is neglected. We model the magnetization subsystem  by the Hamiltonian 
\begin{align}\label{eq:E}
E=S\int\left[-A\,\vec{m}\cdot\vec{\nabla}^2\vec{m}-K(\vec{m}\cdot\vec{e}_{\textsc{t}})^2\right]\mathrm{d}s,
\end{align}
\end{subequations}
which takes into account only two magnetic interactions, namely: the isotropic exchange with the stiffness $A$ (the first integrand) and the easy-tangential anisotropy with the coefficient $K>0$ (the second integrand). Here $S$ denotes the cross-section area, which is assumed constant along the wire. For the wires with the circular or square cross-sections the anisotropy constant reads $K=\pi M_s^2+K_{\rm a}$, where the first term comes from the magnetostatic contribution \cite{Slastikov12} and the second one represents the magnetocrystalline anisotropy. For the magnetically soft wires one has $K_{\rm a}=0$. The model \eqref{eq:Model} determines the system and time- and length-scales presented by the frequency of the uniform ferromagnetic resonance for a straight wire $\omega_0=2K\gamma_0/M_s$, and magnetic length $\ell=\sqrt{A/K}$, respectively. The latter determines typical size of the magnetization nonuniformities, e.g. width of a domain wall. The model \eqref{eq:E} presumes that the magnetization is uniform within the wire cross-section: $\vec{m}=\vec{m}(s,t)$. This assumption is valid for the case $h_{\rm max}\lessapprox\ell$. The latter is the magnetic condition for the `thin wire' approximation. In the following we use the dimensionless time $\tau=t\omega_0$ and length $\xi=s/\ell$. This makes an analysis independent on the material parameters. We utilize the constraint $|\vec{m}|=1$ by means of the angular parametrization $\vec{m}=\sin\theta\left(\cos\phi\vec{e}_\textsc{t}+\sin\phi\vec{e}_\textsc{n}\right)+\cos\theta\vec{e}_\textsc{b}$. For a planar wire the minimization of energy \eqref{eq:E} leads to the static equilibrium state $\theta=\Theta=\pi/2$ and $\phi=\Phi(\xi)$, where the inclination angle $\Phi$ is determined by the driven pendulum equation \cite{Sheka15,Yershov15b}
\begin{equation}\label{eq:Phi}
\Phi''-\sin\Phi\cos\Phi=-\varkappa'.
\end{equation}
Here and below prime denotes the derivative with respect to $\xi$, and $\varkappa=\kappa\ell$ is the dimensionless curvature. The emergent geometry-induced DMI breaks the spatial symmetry of the solution: its contribution is determined by the gradient of the curvature in Eq.~\eqref{eq:Phi}, the only geometrical parameter which influence the magnetization texture of any planar wire.

\begin{figure}
	\centering
\includegraphics[width=0.95\textwidth]{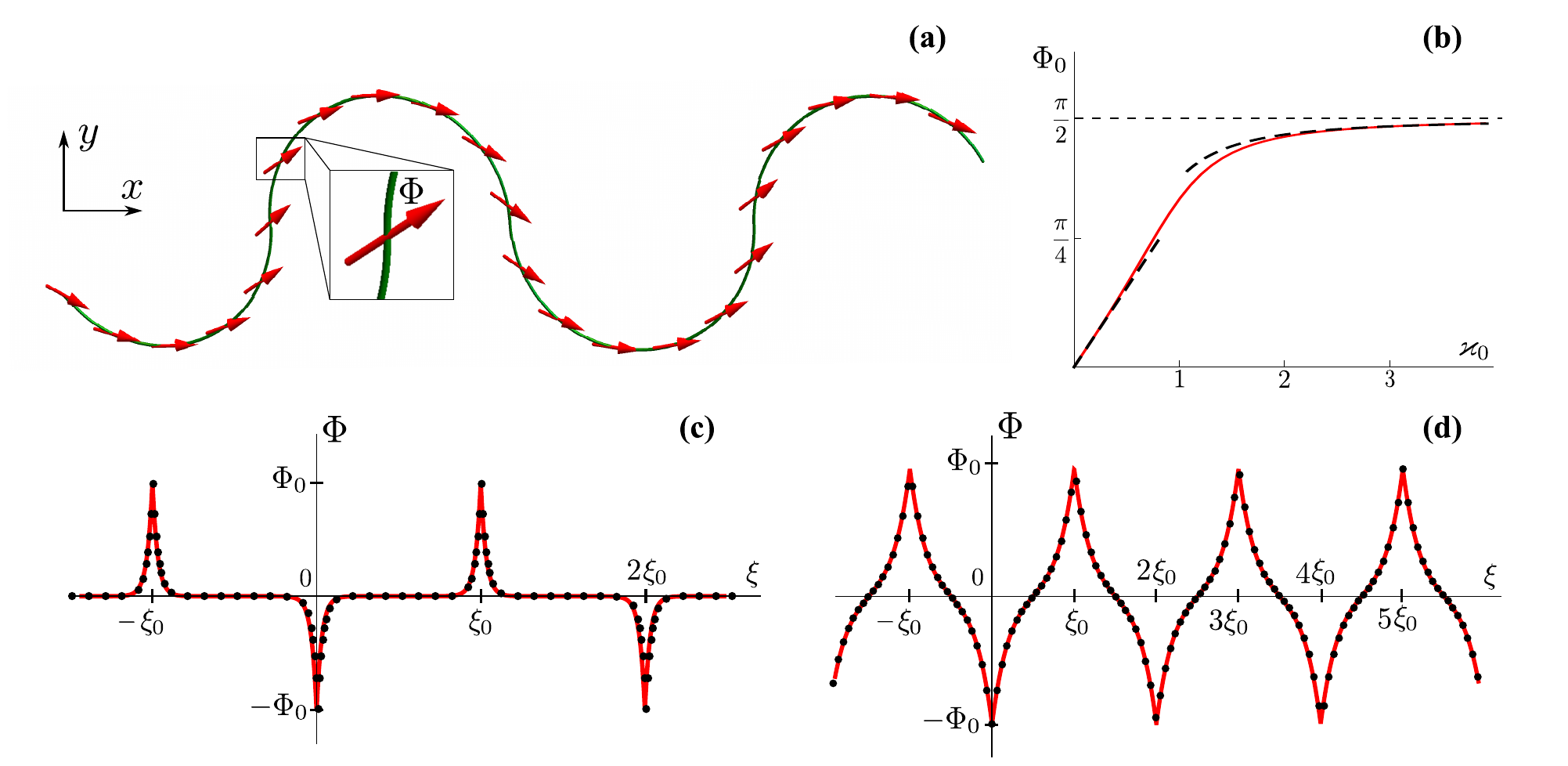}
	\caption{\textbf{Equilibrium state of magnetization:} 
(a) Spatial distribution of magnetization (red arrows) along the wire; the inclination angle $\Phi$ is shown in the inset. The analytical solution \eqref{eq:Phi-sol} (line) is compared to the result of numerical minimization of energy \eqref{eq:E} (dots) for the cases $\varkappa_0=0.1$ (c) and $\varkappa_0=0.6$ (d). The maximal amplitude $\Phi_0$ of the magnetization deviation from the tangential direction is plotted as a function of $\varkappa_0$ in the inset (b): the dashed lines corresponds to asymptotes $\Phi_0\approx\varkappa_0$ and $\Phi_0\approx\frac{\pi}{2} \left(1-\frac{1}{4 \varkappa_0^2}\right)$ valid for the cases $\varkappa_0\ll1$ and $\varkappa_0\gg1$, respectively. }\label{fig2}
\end{figure}

In the following we consider a meander shaped wire in form of the periodically repeated semicircles of radii $R$, see Fig.~\ref{fig1}. The spatial distribution of the curvature of such a wire is the square-wave function with period $2\pi R/\ell$ as shown in  Figs.~\ref{fig1}(c)-(d), see details in Appendix \ref{app:general}. In this case the driving in Eq.~\eqref{eq:Phi} is determined by
\begin{subequations} \label{eq:kappa'-and-Phi}
\begin{equation} \label{eq:kappa'}
\varkappa'=-2\varkappa_0\sum\limits_{n=-\infty}^{\infty}[\delta\left(\xi-2n\xi_0 \right)-\delta\left(\xi-(2n+1)\xi_0\right)],
\end{equation}	
where $\varkappa_0=\ell/R$ is the curvature amplitude, $\xi_0=\pi R/\ell$ is half period of the curvature and $\delta(\bullet)$ is the Dirac delta \cite{NIST10}. The corresponding solution of Eq.~\eqref{eq:Phi} reads
\begin{equation}\label{eq:Phi-sol}
\Phi(\xi)=(-1)^{\lambda} \text{am}\left(\frac{\xi-\xi_0(\lambda + 1/2)}{k},\ ik\right),\qquad \lambda = \left\lfloor\frac{\xi}{\xi_0}\right\rfloor,\quad k=\frac{1}{\sqrt{\varkappa_0^2-\sin^2\Phi_0}},
\end{equation}
\end{subequations}
where $\text{am}(\bullet,\bullet)$ is Jacobi amplitude and $\lfloor \bullet\rfloor$ defines the integer part of $\bullet$ \cite{NIST10}. Constant $\Phi_0=|\Phi(n\xi_0)|$ is maximal value of the function $\Phi(\xi)$, it is determined by the equation $2k F(\Phi_0,ik)=\xi_0$, where $F(\bullet,\bullet)$ is elliptic integral of the first kind and the modulus $k=k(\Phi_0)$ is defined in \eqref{eq:Phi-sol}. Note that the curvature amplitude $\varkappa_0$ is the only parameter which controls the system.

The equilibrium solution \eqref{eq:Phi-sol} is illustrated in Fig.~\ref{fig2}. Magnetization of the equilibrium state always lies within the plane of the wire, but it is not tangential to it. The maximal deviation $\Phi_0$ from the tangential direction takes place in points of junction of two semicircles, this is because of the curvature jump. In the limit case of small curvature $\varkappa_0\ll1$ one has $R\gg\ell$ and due to the easy-tangential anisotropy the wire is magnetized practically tangentially except the junction points, where the magnetization demonstrates the small deviations of amplitude $\Phi_0\approx\varkappa_0$, see Fig.~\ref{fig2}(b, c). In the intermediate case $\varkappa_0\approx1$ the curvature radius is close to the magnetic length ($R\approx\ell$). This means the equilibrium state is determined by strong competition between the anisotropy and exchange interactions, see Fig.~\ref{fig2}(d). The opposite limit case $\varkappa_0\gg1$ corresponds to the regime $R\ll\ell$, when the dominating exchange interaction results in practically uniform magnetization aligned with $x$-axis. This corresponds to the maximal deviation amplitude $\Phi_0\lessapprox\pi/2$, see Fig.~\ref{fig2}(b).

%
%
\section{Band structure}
\label{sec:BS}

\begin{figure}[h]
	\includegraphics[width=\columnwidth]{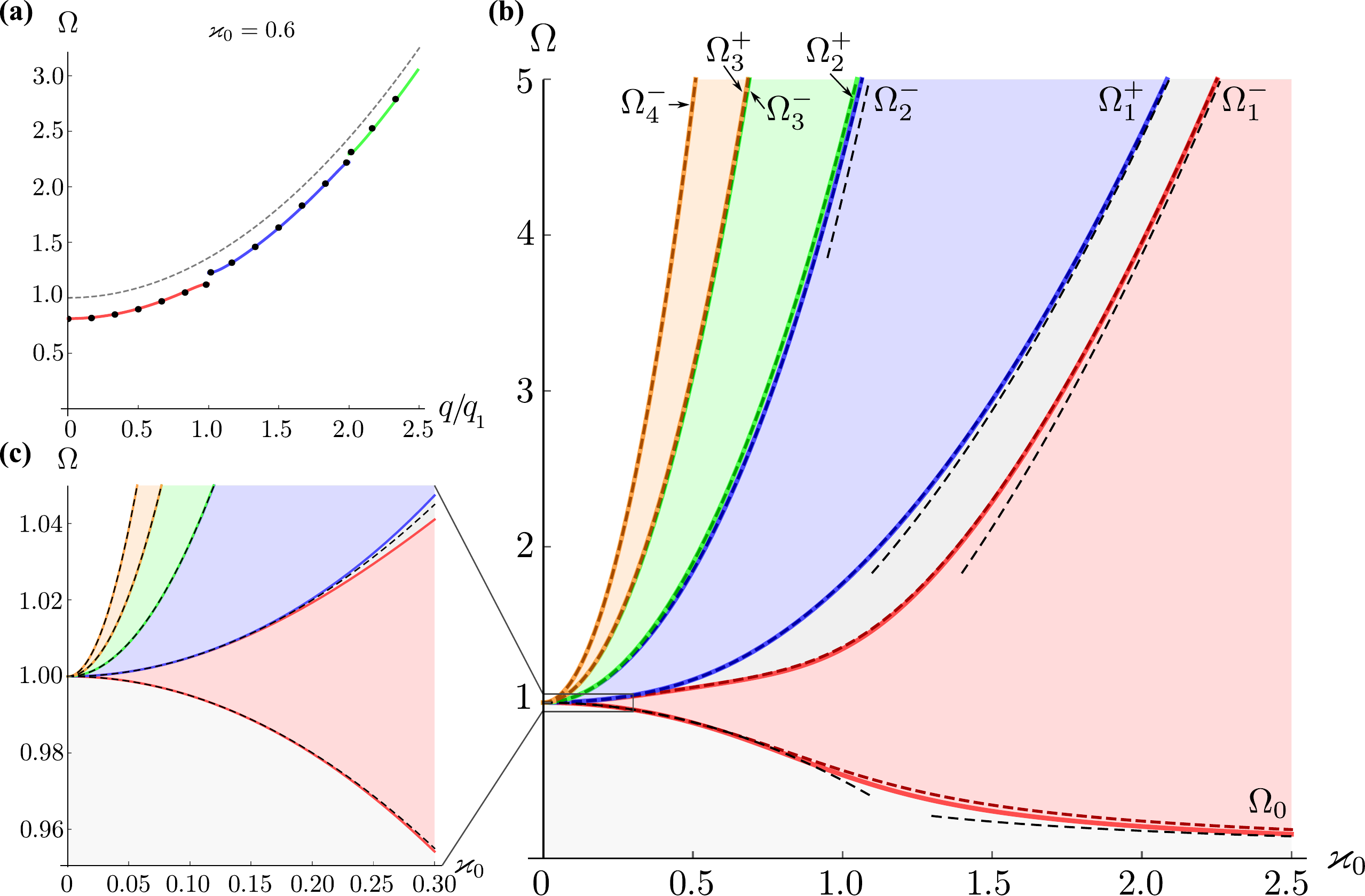}
	\caption{\textbf{Band structure.} (a) Dispersion relation of Eq.~\eqref{eq:gse} for $\varkappa_0 = 0.6$ is obtained both by the numerical solution of the EVP \eqref{eq:BdG-k} (solid lines) and by means of spin-lattice simulations based on the Landau--Lifshitz equation \eqref{eq:B.1.LL} (markers). The dashed line shows the dispersion relation of a straight wire ($\varkappa_0=0$). The wave-vector is normalized by $q_1=\varkappa_0$ corresponding to the edge of the first Brillouin zone. (b) The band structure as function of $\varkappa_0$. The first four bands of the magnon conductance are shown by different colours and gaps are shown by grey shading. Solid and dashed thick lines correspond to the exact numerical solution explained in Appendix~\ref{sbsec:numerics} and to the approximations \eqref{eq:Omega_nu}-\eqref{eq:Omega0}, respectively. The thin dashed lines corresponds to the analytical asymptotes listed in Table~\ref{tbl:asympt}. The inset (c) shows detail of the band structure for small curvatures, the notations are the same as in (b). }\label{fig:BS}
\end{figure}

Here we consider linear dynamics of magnons excited over the equilibrium state described in the previous section. The technique which we use here was previously applied for a number of one-dimensional curvilinear magnetic problems: finding the magnon spectrum for the ring- \cite{Sheka15} and helix-shaped \cite{Sheka15c} wires, localization and scattering of magnon waves on a localized curvilinear defect \cite{Gaididei18a}.

It is convenient to describe the linear magnetization dynamics by means of the complex valued function $\psi=\vartheta+i\varphi$ constructed from deviations from the equilibrium state: $\theta~=~\Theta+\vartheta$ and $\phi=\Phi+\varphi/\sin\Theta$. In this case, Landau--Lifshitz equation linearised in vicinity of the equilibrium state $(\Theta,\Phi)$ has the form of generalized Schr{\"o}dinger equation \cite{Sheka04,Gaididei18a}
\begin{equation}\label{eq:gse}
-i\dot{\psi}=\mathrm{H}\psi+W\psi^*,\qquad \mathrm{H}=-\partial_{\xi\xi}+1+V,
\end{equation}
for details see Appendix~\ref{app:lin}. Here and below the overdot denotes the derivative with respect to the dimensionless time $\tau$. In general case of an arbitrary planar wire the potentials $V$ and $W$ have the form~\cite{Gaididei18a}
\begin{equation}\label{eq:VW}
V=-\frac12\left[3\sin^2\Phi+(\Phi'+\varkappa)^2\right],\qquad W=\frac12\left[\sin^2\Phi-(\Phi'+\varkappa)^2\right],
\end{equation}
where $\Phi$ is determined by Eq.~\eqref{eq:Phi}. Application of Bogoliubov transformation $\psi=u(\xi) e^{i\Omega\tau}+v^*(\xi)e^{-i\Omega\tau}$ to Eq.~\eqref{eq:gse} enables us to formulate an eigenvalue problem (EVP) for a Hamiltonian of Bogoliubov--de~Gennes type
\begin{equation} \label{eq:BdG}
\mathbb{H}|\vec{\Psi}\rangle=\Omega|\vec{\Psi}\rangle,\qquad \mathbb{H}=\begin{Vmatrix}
\mathrm{H} & W\\
-W & -\mathrm{H}
\end{Vmatrix},\quad |\vec{\Psi}\rangle=\begin{Vmatrix}
u\\v
\end{Vmatrix}.
\end{equation}

Equation \eqref{eq:BdG} is the general formulation of the spectral problem for a planar curvilinear wire described by the model \eqref{eq:Model}. Generally, it can be solved numerically if the potentials $V(\xi)$ and $W(\xi)$ are known functions. Here we study a specific case, when the potentials~\eqref{eq:VW} are periodic: $V(\xi)=V(\xi+\xi_0)$ and $W(\xi)=W(\xi+\xi_0)$, see Fig.~\ref{fig:VW}. The periodic potential in a quantum-mechanical Schr{\"o}dinger equation always produces the band structure. The band structure of the 1D system can be expressed quite simply in terms of the properties of scattering in the presence of a single-barrier potential \cite{Ashcroft76}. Similar arguments can be used for the EVP \eqref{eq:BdG}, which results in the band structure with allowed frequency $\Omega(q)~=~\sqrt{\left({\mathcal{K}(q)}^2 +\mathcal{v} \right)^2-\mathcal{w}^2}-1$ and the quasi-wave vector $q$ related by
\begin{equation} \label{eq:band-equation}
\frac{\cos\left(\mathcal{K}\xi_0 + \delta\right)}{T} = \cos q\xi_0,
\end{equation}
where $\mathcal{v} = \max V(\xi)$ and $\mathcal{w} = \min W(\xi)$. Using such an approach the band structure is determined by the scattering data for a single potential: the transmission coefficient $T(\mathcal{K})$ and the scattering phase shift $\delta(\mathcal{K})$, see Appendix \ref{app:Ashcroft} for details. However even the scattering problem can be solved exactly only for few model potentials. Therefore below we also use another method.

\begin{figure}
	\includegraphics[width=\columnwidth]{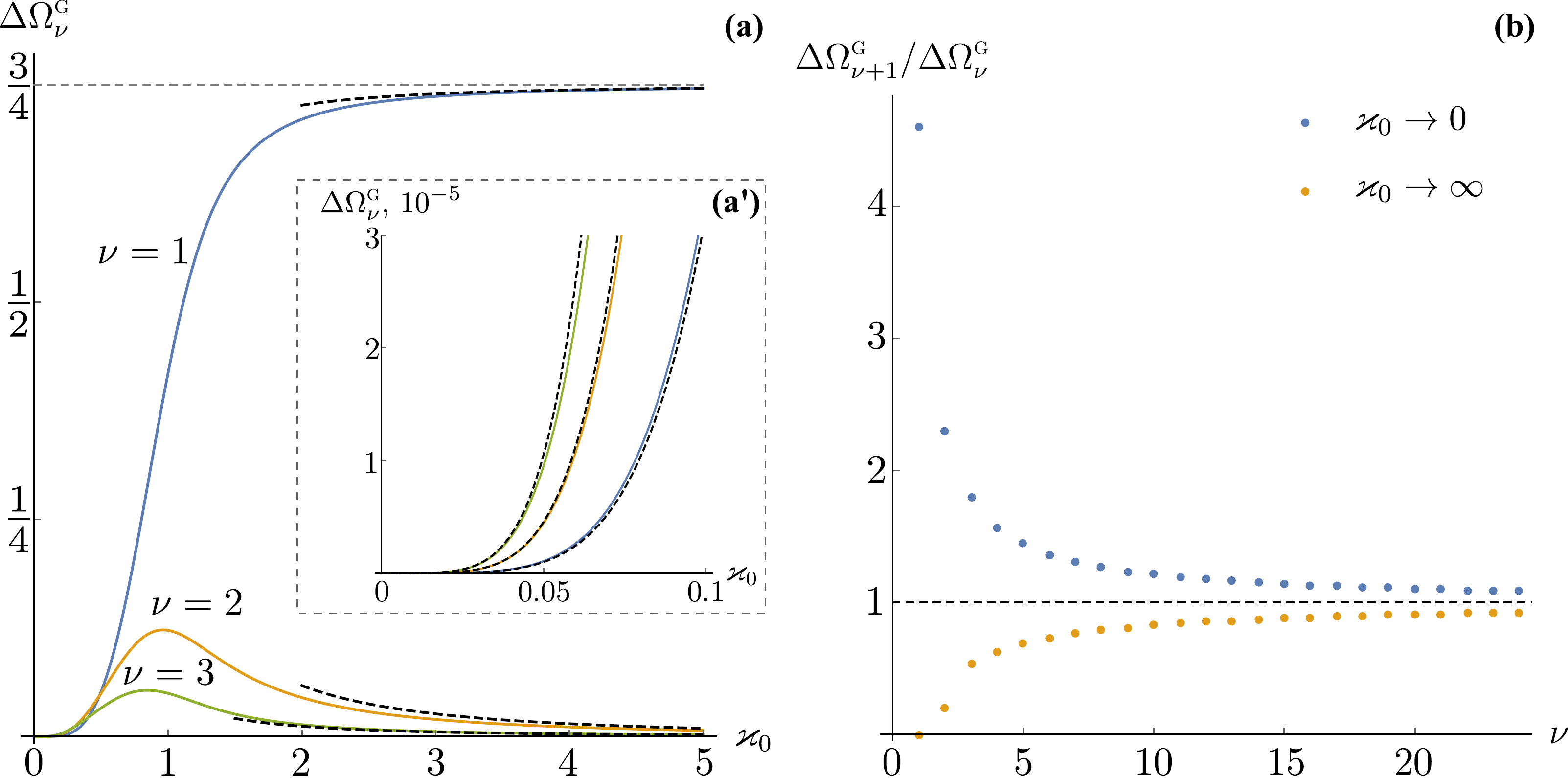}
	\caption{\textbf{Widths of the band gaps:} (a) Exact values of the first three band gaps (solid lines) are compared to the analytical asymptotes listed in Table~\ref{tbl:asympt} (dashed line). The inset (a$'$) shows vicinity of the origin. (b) Ratio of widths of two neighbouring band gaps for opposite limit cases of the vanishing and infinite curvatures.}\label{fig:Fig4}
\end{figure}

Using the periodical properties of potentials, it is convenient to consider the EVP \eqref{eq:BdG} in wave-vector space. For this case one can obtain the numerical solution for the band structure, see Appendix \ref{sbsec:numerics}, as well as analytical asymptotes for the limit cases of small and  large curvatures, see Appendix \ref{sbsct:analyt}. The numerical solutions are shown in Fig.~\ref{fig:BS} by solid lines. In contrast to the straight wire, the dispersion relation for a given curvature $\varkappa_0>0$ demonstrates the frequency reduction and appearance of the band structure, see Fig.~\ref{fig:BS}(a). This is in a good agreement with the direct numerical simulation of the Landau--Lifshitz equation \eqref{eq:LL}. Evolution of the band structure with $\varkappa_0$ is demonstrated in Figs.~\ref{fig:BS}(b),(c).
Generalizing the two-component approximation \cite{Kittel05} for the case of the generalized Schr{\"o}dinger equation \eqref{eq:gse}, we approximate the band gap edges as 
\begin{equation}\label{eq:Omega_nu}
\Omega_\nu^\pm\approx\sqrt{\left(q_\nu^2+1+V_0\mp V_\nu\right)^2-\left(W_0\mp W_\nu\right)^2},\quad \nu \in \mathbb{N}.
\end{equation}
Here $\Omega_\nu^+$ and $\Omega_\nu^-$ correspond to the top and bottom edges of the $\nu$-th gap, respectively. Wave-vector $q_\nu=\nu\varkappa_0$ corresponds to the edge of the $\nu$-th Brillouin zone. Coefficients $V_n$, $W_n$ with $n\in \mathbb{Z}_+$ denote Fourier components \eqref{eq:VW-F-inv} of the potentials. The approximation \eqref{eq:Omega_nu} is shown in Fig.~\ref{fig:BS}(b) by thick dashed lines. The maximal relative deviation from the exact solution (solid lines) is $0.89\%$. Substituting the asymptotic approximations \eqref{eq:VW-Four-ksmall} and \eqref{eq:VW-Four-klarge} for the Fourier coefficients into \eqref{eq:Omega_nu} we obtain the asymptotic behaviour of the gap edges for the limit cases of small and large curvatures, see Table~\ref{tbl:asympt}. For the case $\varkappa_0\ll1$ one has $\Delta\Omega_\nu^{\textsc{g}}\propto\varkappa_0^5$, where $\Delta\Omega_\nu^{\textsc{g}}=\Omega_\nu^+-\Omega_\nu^-$ is the width of the $\nu$-th gap. In the opposite case all the gaps are closed by the law $\Delta\Omega_{\nu\ge2}^{\textsc{g}}\propto\varkappa_0^{-2}$, except the first one, whose width goes to the constant value $\Delta\Omega_{1}^{\textsc{g}}\approx3/4$. 
Numerically found values of the first three band gaps are compared to the analytical asymptotes listed in Table~\ref{tbl:asympt} in Fig.~\ref{fig:Fig4}(a). Interestingly that $\lim\limits_{\nu\to\infty}\Delta\Omega_{\nu+1}/\Delta\Omega_{\nu}=1$ for both cases of the vanishing and infinite curvatures, see Fig.~\ref{fig:Fig4}(b).

The size of the main gap $\Omega_{0}$ can be approximated by means of the one-component approximation
\begin{equation}\label{eq:Omega0}
\Omega_{0}\approx\sqrt{(1+V_0)^2-W_0^2}.
\end{equation}
The simple formula \eqref{eq:Omega0} demonstrates a good agreement with the numerical solution, see Fig.~\ref{fig:BS}(b), therefore it is useful for practical estimations for the main gap width. However, it results in the wrong asymptotic for the limit case $\varkappa_0\gg1$. Applying more exact three-component approximation \eqref{eq:HW-3} we obtain $\Omega_0\approx1-\varkappa_0^2/2$ for $\varkappa_0\ll1$ and $\Omega_0\approx1/(2\sqrt2\varkappa_0)$ for $\varkappa_0\gg1$. The asymptotic closing of the main gap for the large curvatures has the following physical sense. For large curvatures the curvature radius is much smaller than the magnetic length. In this case the wire magnetization is almost uniform and oriented along $x$-axis, see Fig.~\ref{fig2}(c). On the other hand, the anisotropy axis frequently changes its direction within the $xy$-plane when moving along $x$-axis. Thus, in the long-wave approximation, such a wire is physically equivalent to the uniformly magnetized wire with easy-plane anisotropy, which is gapless \cite{Akhiezer68}.

Let us consider the width of the $\nu$-th conductance band $\Delta\Omega_\nu^{\textsc{b}}=\Omega_\nu^--\Omega_{\nu-1}^+$.\footnote{Here we assume that $\Omega_{0}^+=\Omega_{0}^-=\Omega_{0}$.} Using the obtained asymptotic behaviour one can show that $\Delta\Omega_\nu^{\textsc{b}}\approx(2\nu-1)\varkappa_0^2$ for both limit cases $\varkappa_0\ll1$ and $\varkappa_0\gg1$.

%
%

%
%
\section{Conclusion}

We demonstrate that the periodical deformation of an uniaxial ferromagnetic wire results in appearance of the band structure in the magnon spectrum. As a case study we consider a meander-shaped planar wire composed of the semicircles of radius $R$. The normalized curvature $\varkappa_0=\ell/R$ is the only parameter which controls properties of the magnon spectrum. For small curvature $\varkappa_0\ll1$ widths of all band gaps are $\Delta\Omega_{\nu}^{\textsc{g}}\propto\varkappa_0^5$ where $\nu\in \mathbb{N}$. In the opposite case $\varkappa_0\gg1$ all the gaps closes by the law $\Delta\Omega_{\nu\ge2}^{\textsc{g}}\propto\varkappa_0^{-2}$, except the first one, whose width goes to the constant value $\Delta\Omega_{1}^{\textsc{g}}\approx3/4$ (in units of the frequency of the uniform ferromagnetic resonance for a straight wire). The main gap shrinks as $\Omega_0\approx1-\varkappa_0^2/2$ for small curvatures and it asymptotically closes $\Omega_0\approx1/(2\sqrt2\varkappa_0)$ for large curvatures. In the latter case the curvilinear wire effectively behaves as an easy-plane magnet.

The curvature induced band structure is a direct physical evidence of the curvature effect on magnetic subsystem. Besides the fundamental meaning of this result the considered periodical system can be used in as a filter element in magnon spintronics applications.

%
%
\section*{Acknowledgements}
The authors thank Denys Makarov (HZDR) for helpful discussions and Ulrike Nitzsche (IFW) for the technical support.


\paragraph{Funding information}
We acknowledge support from the UKRATOP project funded by the German Federal Ministry of Education and Research, Grant No. 01DK18002 and Leonhard Euler Programm funded by DAAD (Projekt-ID: 57430566). O.V.P. acknowledges the support from DAAD (Grant No. 91530902). O.V.P. and D.D.S. thank Helmholtz-Zentrum Dresden-Rossendorf e.V. (HZDR), where part of this work was performed, for their kind hospitality, and acknowledge the support from the Alexander von Humboldt Foundation (Research Group Linkage Programme). This work was partially supported by Taras Shevchenko National University of Kyiv (Projects 19BF052-01 and 18BF052-01M) and by National Academy of Sciences of Ukraine, Project No. 0116U003192.

\begin{appendix}

\numberwithin{equation}{section}

\section{Equilibrium state and linearised equations of motion}
\label{app:lin}

\subsection{General results}
\label{app:general}

The model \eqref{eq:E} substituted into the angular representation of Landau--Lifshitz equations $\sin\theta\partial_t\phi=\frac{\gamma_0}{M_s}\frac{\delta E}{\delta\theta}$, $\sin\theta\partial_t\theta=-\frac{\gamma_0}{M_s}\frac{\delta E}{\delta\phi}$ results in
\begin{subequations}\label{eq:LL-ang}
\begin{align}
\label{eq:LL-theta}&\sin\theta\dot{\phi}=-\theta''+\sin\theta\cos\theta\left[(\phi'+\varkappa)^2-\cos^2\phi\right],\\
\label{eq:LL-phi}&\sin\theta\dot{\theta}=\left[\sin^2\theta(\phi'+\varkappa)\right]'-\sin^2\theta\cos\phi\sin\phi.
\end{align}
\end{subequations}
One can see that the equation \eqref{eq:LL-theta} has static solution $\theta=\Theta$, where $\Theta=0$ or $\Theta=\pi/2$ corresponding to maximum or minimum of the energy \eqref{eq:E}, respectively. In the following we focus on the static ground state solution $\Theta=\pi/2$. For this case the static form of Eq.~\eqref{eq:LL-phi} is transformed into \eqref{eq:Phi} which determines the ground state $\phi=\Phi(\xi)$. 

Here we consider a meander shaped wire with the periodically repeated semicircles of radii $R$, see Fig.~\ref{fig1}. The curvature of this wire is the square-wave function with period $2\xi_0$:	
\begin{equation} \label{eq:kappa}
\varkappa(\xi)=\varkappa_0 \left(-1\right)^{\left\lfloor \xi/\xi_0 \right\rfloor+1}\!\!.
\end{equation}

For the considered geometry Eq.~\eqref{eq:Phi} is equivalent to the corresponding homogeneous equation supplemented by the derivative jumps in the junction points $\xi=n\xi_0$:
\begin{equation}\label{eq:Phi-det}
\Phi''-\sin\Phi\cos\Phi=0,\qquad [\Phi']_{2n\xi_0}=2\varkappa_0,\quad [\Phi']_{(2n+1)\xi_0}=-2\varkappa_0,
\end{equation}
where $[\Phi']_{\xi}=\Phi'(\xi+0)-\Phi'(\xi-0)$ and $n\in\mathbb{Z}$. The opposite signs of the derivative jumps in the neighbouring junction points enable us to assume that the solution is a periodical function $\Phi(\xi)=\Phi(\xi+2\xi_0)$ reaching its minimum and maximum values in the junction points, namely $\Phi(2n\xi_0)=-\Phi_0$ and  $\Phi((2n+1)\xi_0)=\Phi_0$. Equation \eqref{eq:Phi-det} has the first integral
\begin{equation}\label{eq:Phi-first}
\Phi'=\pm\frac{1}{k}\sqrt{1+k^2\sin^2\Phi},\qquad k=\frac{1}{\sqrt{\varkappa_0^2-\sin^2\Phi_0}},
\end{equation}
where the integration constant $k$ is determined from the derivative jumps  in \eqref{eq:Phi-det}. Note that $\Phi'>0$ and $\Phi'<0$ for the ranges where $\varkappa=-\varkappa_0$ and $\varkappa=\varkappa_0$ respectively. Thus, for all $\xi$ one has
\begin{equation}\label{eq:Phi-kappa}
(\Phi'+\varkappa)^2=\left(\frac{1}{k}\sqrt{1+k^2\sin^2\Phi}-\varkappa_0\right)^2.
\end{equation}
Integration of Eq.~\eqref{eq:Phi-first} on the interval $\xi\in[0,\xi_0]$ results in the solution 
\begin{equation} \label{eq:Phi-1per}
\Phi(\xi)=\text{am}\left(\frac{\xi-\xi_0/2}{k(\Phi_0)},\ ik(\Phi_0)\right).
\end{equation}
The unknown constant $\Phi_0$ can be found from the condition $\Phi(0)=-\Phi_0$ and $\Phi(\xi_0)=\Phi_0$ which is equivalent to the equation $k(\Phi_0)F(\Phi_0,\ ik(\Phi_0))=\xi_0/2$. Generalization of the half-period solution \eqref{eq:Phi-1per} for the all domain $\xi\in\mathbb{R}$ is presented by \eqref{eq:Phi-sol}. Asymptotic approximations for $\Phi_0$ and $\Phi(\xi)$ for the cases of small and large curvatures are listed in Table.~\ref{tbl:asympt}.

\begin{figure}
	\includegraphics[width=\textwidth]{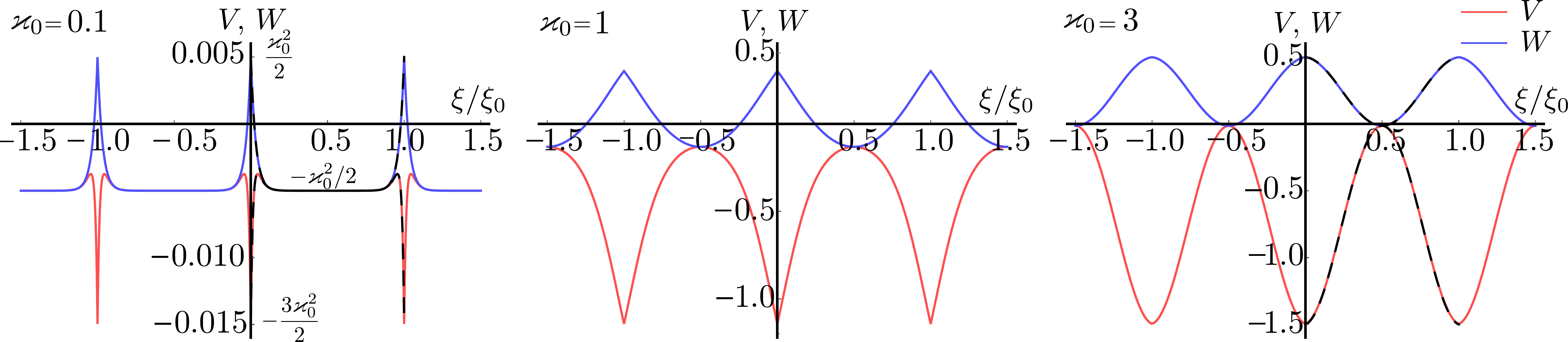}
	\caption{\textbf{Magnon potentials} \eqref{eq:pots} are shown by solid lines for different $\varkappa_0$. Dashed lines correspond to asymptotes \eqref{eq:VW-small-k} and \eqref{eq:VW-large-k} for $\varkappa_0=0.1$ and $\varkappa_0=3$, respectively.   }\label{fig:VW}
\end{figure}

Now we introduce the small deviations from the ground state as described in Sec.~\ref{sec:BS}. The corresponding linearised Landau--Lifshitz equations have form
\begin{equation}\label{eq:LL-lin}
\begin{split}
&\dot{\varphi}=-\vartheta''-\vartheta\left[(\Phi'+\varkappa)^2-\cos^2\Phi\right],\\
&\dot{\vartheta}=\varphi''-\varphi\cos2\Phi.
\end{split}
\end{equation}
Introducing the complex valued function $\psi=\vartheta+i\varphi$ we obtain the generalized Schr{\"o}dinger equation \eqref{eq:gse} with potentials \eqref{eq:VW}. Using \eqref{eq:Phi-kappa} one presents the potentials in form 
\begin{equation} \label{eq:pots}
V=W-2\sin^2\Phi,\qquad 
W=\frac{\varkappa_0}{k}\sqrt{1+k^2\sin^2\Phi}-\frac{1}{2}\left(\frac{1}{k^2}+\varkappa_0^2\right),
\end{equation}
where $\Phi$ is determined in \eqref{eq:Phi-sol}. Potentials \eqref{eq:pots} for different $\varkappa_0$ are shown in Fig.~\ref{fig:VW}. Note that $V(\xi)=V(\xi+\xi_0)$ and $W(\xi)=W(\xi+\xi_0)$ have twice as small period as $\Phi(\xi)$ has, this is because the dependence on $\xi$ is reduced to the dependence on $\sin^2\Phi(\xi)$ in \eqref{eq:pots}.

Looking for solution of \eqref{eq:gse} in form  $\psi=u(\xi)e^{i\Omega\tau}+v^*(\xi)e^{-i\Omega\tau}$ one obtains the following set of homogeneous equations for functions $u$ and $v$~\cite{Gaididei18a}
\begin{subequations}\label{eq:uv}
\begin{align}
-&u''+(1+V)u+Wv=\Omega u,\\
&v''-(1+V)v-Wu=\Omega v,
\end{align}
\end{subequations}
which can be reformulated as EVP \eqref{eq:BdG}.

\section[Band structure for the Bogoliubov--de~Gennes EVP in 1D]{Band structure for the Bogoliubov--de~Gennes\\ EVP in 1D}
\label{app:Ashcroft}

The general analysis of the band structure for the Bogoliubov--de~Gennes EVP \eqref{eq:BdG} 
can be performed, using an approach developed in Ref.~\cite{Ashcroft76} for the usual Scr\"{o}dinger equation. We consider the periodic potentials $V(\xi)$ and $W(\xi)$ as a superposition of single-barrier potentials $\mathcal{V}$ and $\mathcal{W}$ of width $\xi_0$:
\begin{equation} \label{eq:V-n-W-periodic}
V(\xi) = \sum_{n=-\infty}^\infty \mathcal{V} \left(\xi-n\xi_0\right),\qquad W(\xi) = \sum_{n=-\infty}^\infty \mathcal{W} \left(\xi-n \xi_0\right).
\end{equation}
Let us start with the scattering problem for single-barrier potentials $\mathcal{V}(\xi)$ and $\mathcal{W}(\xi)$, which coincide with $V(\xi)$ and $W(\xi)$ within the interval  $\xi\in[-\xi_0,\xi_0]$,  respectively. Outside the interval we suppose, following \cite{Ashcroft76}, that potentials takes their limit values $\mathcal{v} = \max \mathcal{V}(\xi)$ and $\mathcal{w} = \min \mathcal{W}(\xi)$ when $\xi\notin[-\xi_0,\xi_0]$. Then, general solution of eigenfunctions of the EVP  \eqref{eq:BdG} for single potentials in the range $|\xi|>\xi_0/2$ has the following form, cf.\cite{Ashcroft76}
\begin{equation} \label{eq:u-and-v-general}
\begin{aligned}
u(\xi) &=
\begin{dcases*}
C_1 \left(e^{i  \mathcal{K}\xi} + \mathcal{R}e^{-i \mathcal{K} \xi}\right) + C_2 \mathcal{T}e^{-i  \mathcal{K} \xi} + C_3 e^{\mathcal{k} \xi},  & when $\xi< -\frac{\xi_0}{2}$,\\
C_1 \mathcal{T}e^{-i \mathcal{K} \xi} + C_2 \left(e^{-i  \mathcal{K}\xi} + \mathcal{R}e^{i \mathcal{K} \xi}\right) + C_4 e^{-\mathcal{k} \xi},  & when $\xi>\frac{\xi_0}{2}$,
\end{dcases*}
\\
v(\xi)&=
\begin{dcases*}
C_1 \mathcal{p} \left(e^{i \mathcal{K}\xi} + \mathcal{R}e^{-i\mathcal{K} \xi}\right) + C_2 \mathcal{p} \mathcal{T}e^{-i\mathcal{K} \xi} - \frac{C_3}{\mathcal{p}} e^{\mathcal{k} \xi},  & when $\xi< -\frac{\xi_0}{2}$,\\
C_1 \mathcal{p} \mathcal{T}e^{-i\mathcal{K} \xi} + C_2 \mathcal{p} \left( e^{-i \mathcal{K}\xi} + \mathcal{R}e^{i\mathcal{K} \xi}\right) - \frac{C_4}{\mathcal{p}} e^{-\mathcal{k} \xi},  & when $\xi>\frac{\xi_0}{2}$.
\end{dcases*}
\end{aligned}
\end{equation}
Here we use the following notations:
\begin{equation} \label{eq:k-and-kappa}
\begin{aligned}
\mathcal{K}	&= \sqrt{\varepsilon - \mathcal{v}},&
\mathcal{k} &= \sqrt{\varepsilon + \mathcal{v}},&
\mathcal{p}	&= \frac{1+\Omega-\varepsilon}{\mathcal{w}},&
\varepsilon &=  \sqrt{(1+\Omega)^2 + \mathcal{w}^2}
\end{aligned}
\end{equation}
with transmission and reflection complex-valued coefficients being $\mathcal{T}=Te^{i\delta}$ and $\mathcal{R}=R e^{i\delta_r}$, respectively, $C_j$, $j=\overline{1,4}$ being constants.

Now using the Floquet theory, we conclude that $u$ and $v$ have to satisfy the following conditions:
\begin{equation} \label{eq:Bloch-gen}
u(\xi+\xi_0)=e^{iq\xi_0}u(\xi), \qquad v(\xi+\xi_0)=e^{iq\xi_0}v(\xi),
\end{equation}
which are analogues of the Bloch's theorem for the EVP \eqref{eq:BdG}. By imposing conditions \eqref{eq:Bloch-gen} and similar conditions for $u'$ and $v'$ at $\xi=-\xi_0/2$ to \eqref{eq:u-and-v-general}, we find the band structure of EVP \eqref{eq:BdG}:
\begin{equation} \label{eq:band-equationq}
\tag{\ref{eq:band-equation}$'$}
\frac{\cos\left(\mathcal{K}\xi_0 + \delta\left(\mathcal{K}\right)\right)}{T\left(\mathcal{K}\right)} = \cos q\xi_0.
\end{equation}
The band edges are determined by condition 
\begin{equation} \label{eq:band-edge}
\cos\left(\mathcal{K}\xi_0+ \delta \left(\mathcal{K}\right)\right)=\pm
T\left(\mathcal{K}\right).
\end{equation}

\subsection{Numerical band structure calculation in the wave-vector space}
\label{sbsec:numerics}

\begin{figure}
	\centering
	\includegraphics[width=0.8\textwidth]{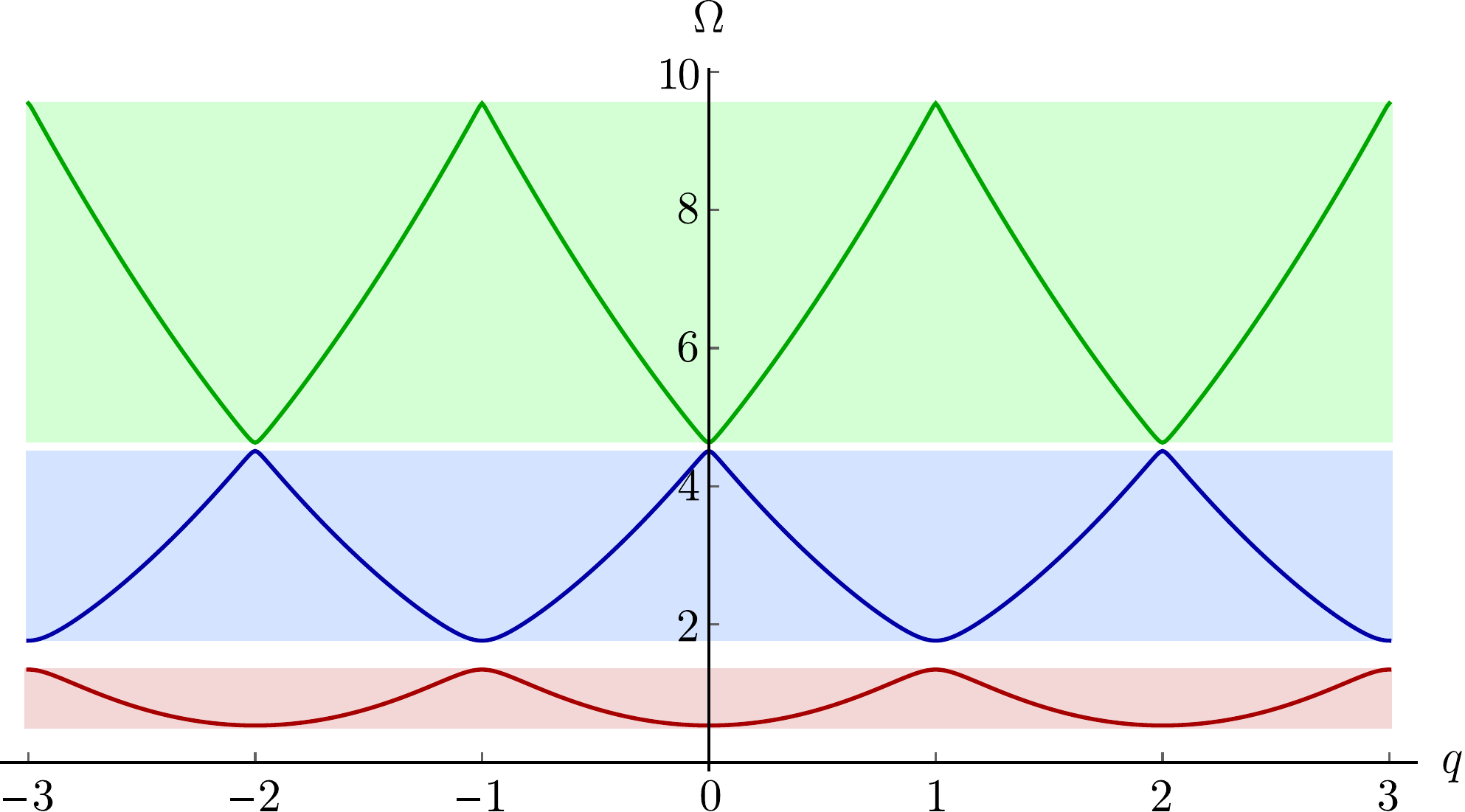}
	\caption{\textbf{Periodical band structure} for the case $\varkappa_0=1$. Three bands are represented by the lowest three eigenvalues of the matrix $\hat{\mathbb{H}}_{q}$ defined in Eq.~\eqref{eq:BdG-k}. The calculation was made for $N_0=10$.}\label{fig:BS-per}
\end{figure}

In the following it is convenient to utilize the periodicity of the potentials $V$ and $W$ and present them in form of the Fourier series
\begin{subequations}
	\begin{align}
\label{eq:VW-F}&V(\xi)=\sum_{n\in\mathbb{Z}}V_ne^{ing_0\xi},\qquad W(\xi)=\sum_{n\in\mathbb{Z}}W_ne^{ing_0\xi},\\
\label{eq:VW-F-inv}&V_n=\frac{1}{\xi_0} \int\limits_{-\xi_0/2}^{\xi_0/2}V(\xi)e^{-ing_0\xi}\mathrm{d}\xi,\qquad W_n=\frac{1}{\xi_0} \int\limits_{-\xi_0/2}^{\xi_0/2}W(\xi)e^{-ing_0\xi}\mathrm{d}\xi,
	\end{align}
\end{subequations}
where $g_0=2\pi/\xi_0=2\varkappa_0$ is vector of the reciprocal lattice. Note that the potentials are even functions $V(\xi)=V(-\xi)$, $W(\xi)=W(-\xi)$. This has two consequences, namely $V_n=V_{-n}$, $W_n=W_{-n}$ and  $V_n=V_{n}^*$, $W_n=W_{n}^*$ for all $n$. Substituting \eqref{eq:VW-F} together with the representation $u(\xi)=\int_{-\infty}^{\infty}\hat{u}(q)e^{iq\xi}\mathrm{d}q$, $v(\xi)=\int_{-\infty}^{\infty}\hat{v}(q)e^{iq\xi}\mathrm{d}q$ into \eqref{eq:uv} and using the orthogonality condition $\int_{-\infty}^{\infty}e^{i\xi(q-q')}\mathrm{d}\xi=2\pi\delta(q-q')$ we obtain the wave-vector representation of \eqref{eq:uv}
\begin{subequations}\label{eq:uv-k}
	\begin{align}
	&(q^2+1)\hat{u}(q)+\sum_{n\in\mathbb{Z}}\left[V_n\hat{u}(q+ng_0)+W_n\hat{v}(q+ng_0)\right]=\Omega_q\hat{u}(q),\\
	-&(q^2+1)\hat{v}(q)-\sum_{n\in\mathbb{Z}}\left[V_n\hat{v}(q+ng_0)+W_n\hat{u}(q+ng_0)\right]=\Omega_q\hat{v}(q).
	\end{align}
\end{subequations}
For each $q$, equations \eqref{eq:uv-k} determine the tuple of numbers $|\hat{\vec{\Psi}}_q\rangle=|\dots,\hat{u}(q+ng_0),\dots,\hat{v}(q+ng_0),\dots\rangle$ which can be interpreted as eigenvector of the following eigenvalue problem
\begin{equation}\label{eq:BdG-k}
\hat{\mathbb{H}}_{q}|\hat{\vec{\Psi}}_q\rangle=\Omega_q|\hat{\vec{\Psi}}_q\rangle,\qquad \hat{\mathbb{H}}_{q}=\begin{Vmatrix}
\hat{\mathrm{H}}_q & \hat{W}\\
-\hat{W} & -\hat{\mathrm{H}}_q
\end{Vmatrix}.
\end{equation}
Here $\hat{\mathbb{H}}_{q}$ is the block-matrix consisting of the blocks $\hat{\mathrm{H}}_q$ and $\hat{\mathrm{W}}$ with elements 
\begin{equation}\label{eq:blocks}
\hat{\mathrm{H}}_q^{nn'}=
\delta_{n,n'}\left[(q+ng_0)^2+1\right]+V_{|n-n'|},\qquad 
\hat{\mathrm{W}}^{nn'}=
W_{|n-n'|},
\end{equation}
where $n,n'\in\mathbb{Z}$. Having the analytical expression \eqref{eq:pots} for the potentials $V$ and $W$ one can easily build matrix $\hat{\mathbb{H}}_{q}$ using \eqref{eq:VW-F-inv} for numerical calculation of the Fourier coefficients $\hat{V}_n$ and $\hat{W}_n$. The periodical band scheme for the first $m$ zones is determined by $m$ lowest eigenvalues of the matrix $\hat{\mathbb{H}}_{q}$, see Fig.~\ref{fig:BS-per}. In this case the size of the matrix $\hat{\mathbb{H}}_{q}$ must be much larger than $m$: $-N_0\le n,n'\le N_0$, where $N_0\gg m$.

\subsection{Analytical approximation}
\label{sbsct:analyt}

In vicinity of the $\nu$-th Brillouin zone ($\nu\in\mathbb{N}$) the EVP \eqref{eq:BdG-k} can be significantly simplified by applying the two-component approximation \cite{Kittel05}. The solution is approximated by the standing wave with $q=q_\nu$, where the wave vector $q_\nu=\nu g_0/2=\nu\varkappa_0$ corresponds to the edge of the $\nu$-th Brillouin zone. This means that in the general set \eqref{eq:uv-k} one saves only terms $\hat{u}(q_\nu)$, $\hat{u}(-q_\nu)$, $\hat{v}(q_\nu)$, and $\hat{v}(-q_\nu)$. For this case, the block matrices $\hat{\mathrm{H}}_q$ and $\hat{W}$ in \eqref{eq:BdG-k} have simple form
\begin{equation}\label{eq:HW-simple}
\hat{\mathrm{H}}_{q_\nu}=\begin{Vmatrix}
q_\nu^2+1+V_0 & V_\nu\\
V_\nu & q_\nu^2+1+V_0
\end{Vmatrix},\qquad \hat{W}=\begin{Vmatrix}
W_0 & W_\nu \\
W_\nu & W_0
\end{Vmatrix}.
\end{equation}
Using the fact that matrices $\hat{\mathrm{H}}_{q_\nu}$ and $\hat{W}$ commute, one obtains from \eqref{eq:BdG-k} the following eigenvalues $\Omega_{q_\nu}=\Omega_\nu^\pm$. 
Here $\Omega_\nu^+$ and $\Omega_\nu^-$ correspond to the upper and bottom edges of the $\nu$-th gap, respectively, they are determined by Eq.~\eqref{eq:Omega_nu}. The value of the main gap $\Omega_0$, which corresponds to the case $\nu=0$ and $q_0=0$, can be roughly approximated by the one component approximation \eqref{eq:Omega_nu}. Although, the latter works well in a wide range of $\varkappa_0$, it does not give the correct asymptotic for $\varkappa_0\gg1$. For this case one should apply the three-component approximation:
\begin{equation}\label{eq:HW-3}
\hat{\mathrm{H}}_{q_0}=\begin{Vmatrix}
g_0^2+1+V_0 & V_1 & 0\\
V_1 & 1+V_0 & V_1 \\
0 & V_1 & g_0^2+1+V_0
\end{Vmatrix},\qquad \hat{W}=\begin{Vmatrix}
W_0 & W_1 & 0 \\
W_1 & W_0 & W_1 \\
0 & W_1 & W_0 
\end{Vmatrix}.
\end{equation}
The straightforward calculation of eigenvalues of the matrix  $\hat{\mathbb{H}}_{q_0}$ and the subsequent substitution of the asymptotic approximations for the Fourier coefficients $V_{0,1}$ and $W_{0,1}$ listed in \eqref{eq:VW-Four-ksmall} and \eqref{eq:VW-Four-klarge}, enable us to estimate $\Omega_0\approx1-\varkappa_0^2/2$ for $\varkappa_0\ll1$ and $\Omega_0\approx1/(2\sqrt2\varkappa_0)$ for $\varkappa_0\gg1$.

\begin{table}
	\centering
	\renewcommand{\arraystretch}{1.2}
	\begin{tabular}{c|c|c|c|c|c}
		\hline\hline
		Quantity & $\varkappa_0\ll1$ & $\varkappa_0\gg1$ & Quantity & $\varkappa_0\ll1$ & $\varkappa_0\gg1$\\ \hline
		$\Omega_0$ &$1-\varkappa_0^2/2$ & $1/(2\sqrt2\varkappa_0)$& & & \\ \hline
		$\Omega_1^-$ & $1+\frac{\varkappa_0^2}{2}$ & $\varkappa_0^2-\frac18$ &$\Omega_1^+$&$1+\frac{\varkappa_0^2}{2}$&$\varkappa_0^2+\frac58$ \\
		 \hline
	$\Delta\Omega_1^{\textsc{g}}$&$\frac{10}{\pi}\varkappa_0^5$&$\frac34-\frac{3}{32\varkappa_0^2}$&$\Delta\Omega_1^{\textsc{b}}$&$\varkappa_0^2$&$\varkappa_0^2$\\
	\hline
	$\Omega_2^-$&$1+\frac{7\varkappa_0^2}{2}$&$4\varkappa_0^2+\frac14$&$\Omega_2^+$&$1+\frac{7\varkappa_0^2}{2}$&$4\varkappa_0^2+\frac14$\\
	\hline
	$\Delta\Omega_2^{\textsc{g}}$&$\frac{46}{\pi}\varkappa_0^5$&$\frac{15}{64\varkappa_0^2}$&$\Delta\Omega_2^{\textsc{b}}$&$3\varkappa_0^2$&$3\varkappa_0^2$\\
	\hline
	$\Omega_3^-$&$1+\frac{17\varkappa_0^2}{2}$&$9\varkappa_0^2+\frac14$&$\Omega_3^+$&$1+\frac{17\varkappa_0^2}{2}$&$9\varkappa_0^2+\frac14$\\
	\hline
	$\Delta\Omega_3^{\textsc{g}}$&$\frac{106}{\pi}\varkappa_0^5$&$\frac{3}{64\varkappa_0^2}$&$\Delta\Omega_3^{\textsc{b}}$&$5\varkappa_0^2$&$5\varkappa_0^2$\\
		\hline\hline
	\end{tabular}
	\caption{\textbf{Asymptotic behaviour of the band boundaries:} Only the leading terms in $\varkappa_0$ are presented. Here $\Omega_0$ is width of the main gap, $\Omega_\nu^-$ and $\Omega_\nu^+$ denote the bottom and the top edges of the $\nu$-th gap with $\nu\in\mathbb{N}$. The notations $\Delta\Omega_\nu^{\textsc{g}}=\Omega_\nu^+-\Omega_\nu^-$ and $\Delta\Omega_\nu^{\textsc{b}}=\Omega_\nu^--\Omega_{\nu-1}^+$ are introduced for the widths of the $\nu$-th gap and the conductance band, respectively (we define $\Omega_{0}^+=\Omega_{0}^-=\Omega_{0}$).}\label{tbl:asympt}
\end{table}

	

\subsubsection{Asymptotic behaviour of the ground state and the magnon potentials }

I.~\textit{Small curvature $\varkappa_0\ll1$}.

The ground state \eqref{eq:Phi-sol} within the interval $\xi\in[0,\xi_0]$: 
\begin{equation}\label{eq:Phi0-small-k}
\Phi(\xi)\approx\varkappa_0\frac{\sinh\left(\xi-\frac{\xi_0}{2}\right)}{\cosh\frac{\xi_0}{2}},\qquad \Phi_0\approx\varkappa_0\tanh\frac{\xi_0}{2}.
\end{equation}

Magnon potentials \eqref{eq:pots} within the interval $\xi\in[0,\xi_0]$:
\begin{subequations}\label{eq:VW-small-k}
\begin{align}
&V(\xi)\approx\varkappa_0^2\left[\frac{\cosh\left(\xi-\frac{\xi_0}{2}\right)}{\cosh\frac{\xi_0}{2}}-2\frac{\sinh^2\left(\xi-\frac{\xi_0}{2}\right)}{\cosh^2\frac{\xi_0}{2}}-\frac{1}{2\cosh^2\frac{\xi_0}{2}}-\frac12\right],\\
&W(\xi)\approx\varkappa_0^2\left[\frac{\cosh\left(\xi-\frac{\xi_0}{2}\right)}{\cosh\frac{\xi_0}{2}}-\frac{1}{2\cosh^2\frac{\xi_0}{2}}-\frac12\right],\\
&V(0)\approx-\frac32\varkappa_0^2,\quad W(0)\approx\frac12\varkappa_0^2,\quad V(\xi_0/2)=W(\xi_0/2)\approx-\frac12\varkappa_0^2.
\end{align}
\end{subequations}

Fourier coefficients \eqref{eq:VW-F-inv}
\begin{subequations}\label{eq:VW-Four-ksmall}
	\begin{align}
	&V_0\approx-\frac{\varkappa_0^2}{2}\tanh^2\frac{\xi_0}{2},\quad V_1\approx-\frac{6\pi^2\varkappa_0^2\xi_0\tanh\frac{\xi_0}{2}}{(2\pi^2+\xi_0^2)^2+\pi^2\xi_0^2},\quad V_2\approx-\frac{24\pi^2\varkappa_0^2\xi_0\tanh\frac{\xi_0}{2}}{(8\pi^2+\xi_0^2)^2+4\pi^2\xi_0^2}\\
	&W_0\approx\varkappa_0^2\left[\frac{2}{\xi_0}\tanh\frac{\xi_0}{2}-\frac12\left(1+\frac{1}{\cosh^2\frac{\xi_0}{2}}\right)\right],\quad W_1\approx\frac{2\varkappa_0^2\xi_0\tanh\frac{\xi_0}{2}}{4\pi^2+\xi_0^2},\quad W_2\approx\frac{2\varkappa_0^2\xi_0\tanh\frac{\xi_0}{2}}{16\pi^2+\xi_0^2}.
	\end{align}
\end{subequations}

II.~\textit{Large curvature $\varkappa_0\gg1$}.

The ground state \eqref{eq:Phi-sol} within the interval $\xi\in[0,\xi_0]$:
\begin{equation}\label{eq:Phi0-large-k}
\Phi(\xi)\approx \left(\xi-\frac{\xi_0}{2}\right)\left(\varkappa_0-\frac{1}{4\varkappa_0}\right)+\frac{\sin\left(\pi\frac{\xi}{\xi_0}\right)\cos\left(\pi\frac{\xi}{\xi_0}\right)}{4\varkappa_0^2},\qquad\Phi_0\approx\frac{\pi}{2}\left(1-\frac{1}{4\varkappa_0^2}\right).
\end{equation}

Magnon potentials \eqref{eq:pots} within the interval $\xi\in[0,\xi_0]$: 
\begin{subequations}\label{eq:VW-large-k}
	\begin{align}
	&V(\xi)\approx-\frac32\cos^2\left(\pi\frac{\xi}{\xi_0}\right)-\frac{3 \left(\xi-\frac{\xi_0}{2}\right)\sin\left(2\pi\frac{\xi}{\xi_0}\right)}{8\varkappa_0}+ \frac{\sin\left(\pi\frac{\xi}{\xi_0}\right)\left[3\sin\left(\pi\frac{\xi}{\xi_0}\right)+7\sin\left(3\pi\frac{\xi}{\xi_0}\right)\right]}{32\varkappa_0^2},\\ 
	&W(\xi)\approx\frac12\cos^2\left(\pi\frac{\xi}{\xi_0}\right)+\frac{\left(\xi-\frac{\xi_0}{2}\right)\sin\left(2\pi\frac{\xi}{\xi_0}\right)}{8\varkappa_0}- \frac{\sin\left(\pi\frac{\xi}{\xi_0}\right)\left[5\sin\left(\pi\frac{\xi}{\xi_0}\right)+\sin\left(3\pi\frac{\xi}{\xi_0}\right)\right]}{32\varkappa_0^2},\\
	&V(0)\approx-\frac32,\quad W(0)\approx\frac12,\quad V(\xi_0/2)=W(\xi_0/2)\approx-\frac{1}{8\varkappa_0^2}.
	\end{align}
\end{subequations}

Fourier coefficients \eqref{eq:VW-F-inv}
\begin{subequations}\label{eq:VW-Four-klarge}
	\begin{align}
	&V_0\approx-\frac34\left(1-\frac{5}{16\varkappa_0^2}\right),\quad V_1\approx-\frac38+\frac{5}{64\varkappa_0^2},\quad V_2\approx\frac{-15}{128\varkappa_0^2},\\
	&W_0\approx\frac14\left(1-\frac{9}{16\varkappa_0^2}\right),\quad W_1\approx\frac18+\frac{1}{64\varkappa_0^2},\quad W_2\approx\frac{11}{384\varkappa_0^2}.
	\end{align}
\end{subequations}

%
%
\section{Numerical simulations}\label{sec:simuls}

\subsection{Magnon dispersion}

To verify our results, we numerically simulate the magnetization dynamics of the meander shape chain,~Fig.~\ref{fig1}(a) of the discrete magnetic moments $\bm{m}_i$, with $i =\overline{1,N}$. The magnetization dynamics is determined by the discrete Landau--Lifshitz equations with the relaxation term 

\begin{equation}
\label{eq:B.1.LL}
\frac{2}{\omega_0} \frac{d\bm{m}_i}{dt} = \bm{m}_i \times \frac{\partial \mathcal{E} }{\partial\bm{m}_i} -\alpha \bm{m}_i\times\left[\bm{m}_i \times \frac{\partial \mathcal{E}}{\partial \bm{m}_i} \right] , 
\end{equation}
with $\omega_0=2K\gamma_0/M_s$, $\alpha$ being the damping parameter and $\mathcal{E}$ being the dimensionless energy normalized by $K \Delta s^3$ with $\Delta s$ being the lattice constant. The Hamiltonian of the magnetic wire reads

\begin{equation}
\label{eq:B.1.Hamiltonian}
\mathcal{E} = -2 \frac{\ell^2}{\Delta s^2} \sum_{i=1}^{N-1}\bm{m}_i\cdot\bm{m}_{i+1} -  \sum_{i=1}^{N}\left[\left(\bm{m}_i\cdot\bm{e}_{\textsc{t}_i}\right) ^2 + \bm{b}_i\cdot\bm{m}_i \right] 
\end{equation}
 with $\bm{b}_i$ being a dimensionless external magnetic field, normalized by 
 $K/M_s$. The magnetization dynamics is described by a set of $3N$ ordinary differential equations~\eqref{eq:B.1.LL} for unknown magnetization components $m_i^x(t), m_i^y(t), m_i^z(t)$, $i=\overline{1,N}$. The set of equation~\eqref{eq:B.1.LL} was integrated numerically.

We consider $20$ periods of meander-like shape chain which corresponds to the length $L=2000\Delta s$ and the magnetic length $l = 9.6\Delta s$. During the simulation the discretization step is much smaller than the length of the single half-circle, $\Delta s \ll \pi/\varkappa_0$.

The magnon dispersion relation is carried out in three steps. In the first step, we numerically simulate the magnetization dynamics for the system with Hamiltonian~\eqref{eq:B.1.Hamiltonian} and $\bm{b}_i = 0$ in overdamping regime, $\alpha=0.1$. The tangential distribution is chosen as initial one. The obtained magnetization distribution corresponds to the ground state.
Then the external magnetic field was applied along the shaped wire

\begin{equation}
\label{eq:B.1.Field}
\bm{b}_i =b_0 \bm{e}_{\textsc{b}_i}\cos s_i \frac{q}{\ell}, \quad i=\overline{1,N}
\end{equation} 
where $q$ is the dimensionless wave vector, $s_i = (i-1)\Delta s$ is the position of the magnetization vector $\bm{m}_i$ and the amplitude of the external magnetic field $b_0 =0.005$. 

In the last step, the magnetic field is switched off and the magnetization dynamics is simulated with damping coefficient $\alpha=0.01$. After that the space-time Fourier transform is performed for the binormal magnetization component. The frequency $\Omega$ corresponds to the maximum of the Fourier amplitude and the wave vector $q$ is marked by a dot in Fig.~\ref{fig:BS}.

\subsection{Amplitude-frequency characteristic of the filter}\label{sec:AFC}

We demonstrate that the magnon filter is a possible application for the proposed curvature induced magnonic crystal. We perform spin-lattice simulations of the magnetization dynamics, determined by Landau--Lifshitz equation~\eqref{eq:B.1.LL}. The meander shape chain of the discrete magnetic moments $\bm{m}_i$, with $i =\overline{1,N}$, lies in the plane $x0y$. The full energy of the magnetic system corresponds to the Hamiltonian~\eqref{eq:B.1.Hamiltonian} with $\bm{b}_i=0$.

To form the input signal the first magnetic moment of the system is rotated with the input frequency $\Omega$, $\bm{m}_1 =\hat{\bm{x}}\cos\Omega\tau\sin\beta+\hat{\bm{y}}\cos\beta+\hat{\bm{z}}\sin\Omega\tau\sin \beta $. We consider the 14 periods of the meander-like periodical structure with damping coefficient $\alpha=0.01$. To prevent signals reflection from the free boundary of the system the terminator is placed after the considered magnon crystal. The terminator is composed of 5 periods meander shape chain with $\mathcal{N}$ nodes and has the spatially inhomogeneous damping coefficient, $\alpha_j =\alpha\exp\left( \chi s_j\right) $, $j =\overline{N-\mathcal{N}, N}$ and the coefficient $\chi$ determines the rate of increase of the damping parameter.

The output signal is measured for the last node of the magnonic crystal. Due to non-linear processes, there is a transfer of energy density from the input frequency $\Omega$ to other spectral frequencies. To analyse the amplitude of the output signal on the input frequency the time Fourier transform is performed for one of the magnetization components. 

Simulations are performed for curvature amplitudes $\varkappa_0 = 0.6$ and $\varkappa_0 = 1.2$, see Fig.~\ref{fig1}(d). In the first case, the length of the periodic structure which corresponds to the magnonic crystal is $1400 \Delta s$ and the first magnetic moment is rotating with $\beta=\pi/2$. The length of the terminator is $500 \Delta s$ and $\chi = 0.01$. In the second case, the length of the magnonic crystal is $700 \Delta s$ and the first magnetic moment is rotating with $\beta=\pi/12$. The length of the terminator is $250 \Delta s$ and $\chi = 0.025$.

\subsubsection{Supplemental movie description}\label{sec:movie}
In order to demonstrate the magnetization dynamics of the meander-shaped wire, a supplemental movie is provided. It illustrates the magnetization dynamics obtained by spin-lattice simulations based on the Landau-Lifshitz equation ~\eqref{eq:B.1.LL}, see App.~\ref{sec:AFC} for details.
The video demonstrates the magnetization dynamics for 14 periods of wave-shaped wire with curvature $\varkappa_0 = 1.2$. The terminator was placed after the magnonic crystal to prevent signal reflection. The $m_z$ magnetization component is shown by color. The input signal is formed by rotation of the first magnetic moment of the system with frequencies $\Omega =  1$ and $\Omega =  1.8$ which correspond to the conductive band and band gap, respectively. The input and output signals are measured in the first and last node of the magnonic crystal, as shown in the video. The signal with frequency $\Omega =  1.8$ which corresponds to the band gap is strongly depressed inside the magnonic crystal.

\end{appendix}



\bibliography{periodic}

\nolinenumbers

\end{document}